\documentclass[preprint,tightenlines,aps,floats,nofootinbib]{revtex4}

\usepackage{graphicx}
\usepackage[dvips]{color}
\usepackage{amsmath,amssymb,bbm,mathrsfs}

\newcommand{\notE}{\ \hbox{{$E$}\kern-.60em\hbox{/}}}
\newcommand{\notp}{\ \hbox{{$p$}\kern-.43em\hbox{/}}}
\def\D0{\mbox{D\O}}

\newcommand{\eps}{\epsilon}

\includegraphics{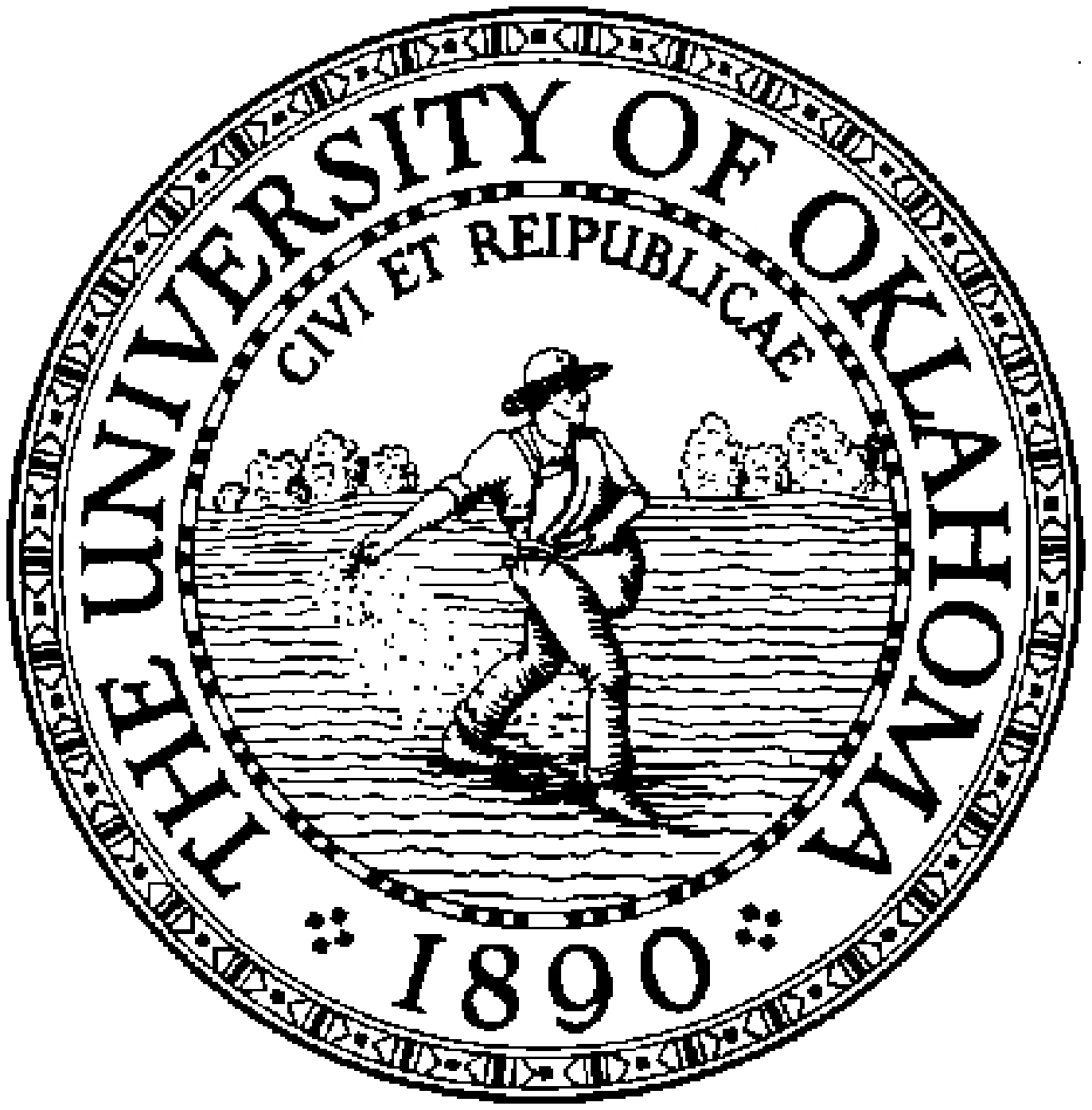}

\preprint{\font\fortssbx=cmssbx10 scaled \magstep2
\hbox to \hsize{
\hskip1.2in 
\hbox{\fortssbx The University of Oklahoma}
\hskip0.2in $\vcenter{
                      \hbox{\bf OUHEP-121122}
                      \hbox{\bf arXiv: [hep-ph]}
                      \hbox{December 2012}}$ }
}

\begin{document}
 
\title{\vspace*{0.7in}
Confirming the LHC Higgs Discovery with $WW$}

\author{
Chung Kao\thanks{Email address: kao@physics.ou.edu} and 
Joshua Sayre\thanks{Email address: sayre@physics.ou.edu}
}

\affiliation{
Homer L. Dodge Department of Physics and Astronomy \\
University of Oklahoma \\ 
Norman, Oklahoma 73019, USA 
\vspace*{.4in}}

\date{\today}

\begin{abstract}

We investigate the prospects of observing a neutral Higgs boson 
decaying into a pair of $W$ bosons (one real and the other virtual), 
followed by the $W$ decays into $qq' \ell\nu$ or $jj\ell\nu$ 
at the CERN Large Hadron Collider (LHC). 
Assuming that the missing transverse energy comes solely from the
neutrino in $W$ decay, we can reconstruct the $W$ masses and then 
the Higgs mass. At the LHC with a center of mass energy ($\sqrt{s}$)
of 8 TeV and an integrated luminosity ($L$) of 25 fb$^{-1}$, 
we can potentially establish a $6\sigma$ signal. 
A $5\sigma$ discovery of $H \to WW^* \to jj\ell\nu$ for $\sqrt{s} = 14$
TeV can be achieved with $L = $ 6 fb$^{-1}$. 
The discovery of $H \to WW$ implies that the recently discovered 
new boson is a CP-even scalar if its spin is zero.
In addition, this channel will provide a good opportunity 
to study the $HWW$ coupling. 

\end{abstract}

%

\maketitle


\color{black}

\section{ Introduction}

Recently, searches for a Standard Model (SM) Higgs boson at both 
the ATLAS and CMS experiments have furnished compelling evidence 
for a new particle consistent with a Higgs boson, having a mass near 
125 GeV~\cite{cmshiggs,atlashiggs}. 
Both collaborations report an apparent excess of events in the $\gamma\gamma$ 
channel and the $ZZ \to 4\ell$ channel. At the same time, searches in the 
channel $WW \to \ell\nu\ell\nu$ have excluded a SM Higgs boson at the 
$95 \%$ confidence level for masses above $129$ GeV and combined
searches exclude from 110 to 122.6 GeV~\cite{atlasllvv, cmsllvv}. 
Results from LEP II preclude the mass region below 
$114$ GeV~\cite{Barate:2003sz}. 
The LHC is now taking data for its 2012 run at a center of mass (CM) 
energy of 8 TeV, after which it is planned to be offline for a year 
before an upgraded run in 2014.

With the detection of this new particle, the Higgs program at the LHC
moves into a new phase of testing to determine the properties of 
the particle in as much detail as possible. For this mass range, 
the $\gamma \gamma$ and $ZZ \to 4l$ channels should continue to
provide the best mass resolution. However, it is worthwhile to
consider all potential channels which can make any significant 
contribution, both to refine our results with additional data and to 
test the consistency of any discovery with the Standard Model or 
its variations.

Towards this end we consider the potential for detection of the Higgs 
decaying to $W^+ W^-$, where one $W$ decays hadronically and one 
leptonically, $H \to WW^* \to qq'\ell\nu$ or $jj\ell\nu$. This channel 
has previously been considered for the Tevatron and the LHC, but 
generally not for such a low Higgs mass~\cite{Han:1998ma,Lykken:2011uv}. 
The ATLAS collaboration has released the results of a search in this 
channel with $4.7 \; \text{fb}^{-1}$ of data at 7 TeV, focusing on
the 300 to 600 GeV mass range, for which they find no significant 
excess, although this channel by itself does not yet exclude the
expected Standard Model cross-section~\cite{atlasjjlv}. 
CMS finds no evidence for a Higgs in the range 170 to 600 GeV and 
excludes a SM Higgs for 230-480 GeV~\cite{cmsjjlv}. 
In this letter we will consider the lower mass range consistent with 
the announced discovery.

The $jj\ell\nu$ signal presents some difficulties in this relatively 
low mass range. First, it is clearly well below the nominal $WW$
production threshold, so that a resonantly produced Higgs boson will 
decay with at least one $W$ far below its mass shell. This means that 
$WW$ is not the leading branching fraction and our signal is smaller 
than it would be at higher masses. On the other hand, as we shall see, 
this far-off-shell case presents some kinematic characteristics which 
can help distinguish it from the backgrounds. 

The second problem is that introducing jets in our signal inevitably 
involves dealing with large QCD backgrounds. As mentioned above, 
$WW \to \ell\nu\ell\nu$ is currently being searched and has already 
yielded strong upper limits on a SM Higgs. This channel has the
advantage of having primarily backgrounds from weak interactions. 
On the other hand, the presence of two neutrinos limits our ability 
to reconstruct the event kinematics. Allowing one of the $W$s to decay 
hadronically means we must contend with the large $Wjj$ background, 
but has the advantage of including only one neutrino in the signal.

Although the single neutrino still presents us with an unmeasured 
momentum, we can determine its components as described in our analysis 
and identify a characteristic mass peak near the physical Higgs mass 
nonetheless. We estimate the rates for the signal and the background with
appropriate cuts and show that this channel ($H \to WW^* \to jj\ell\nu$)
can contribute a $6\sigma$ statistical significance by itself with 
$\sqrt{s} = 8$ TeV and an integrated luminosity ($L$) of 25 fb$^{-1}$ 
for the 2012 running in each experiment (ATLAS or CMS) at the LHC.
We also find that an independent $5\sigma$ discovery of a 125 GeV Higgs
boson in this channel can be achieved for the design CM energy of 
14~TeV with $L = 6 \; \text{fb}^{-1}$.

Additionally, we consider a proposal by Sullivan and Menon to augment 
this channel with the development of c-tagging 
algorithms~\cite{Menon:2010vm}. 
We show that with ideally perfect c-tagging, one could potentially 
increase the significance of the signal in 2012 data to $9.7 \sigma$.
With modest assumptions for c-tagging performance we find 
only marginal improvements to the statistical significance, 
although the ratio of signal to background would be improved.

In Section II, we describe the characteristics of the signal and 
the background for $H \to WW^* \to jj\ell\nu$. 
Section III presents our strategy to reconstruct the Higgs signal for the
final state with one neutrino. Sections IV and V describe details of 
our simulations and acceptance cuts. Promising results are shown 
in Section VI, and prospects with c-tagging are discussed in Section VII. 
Optimistic conclusions are drawn in Section VIII.

\section{Signal and Background Characteristics}

For a typical signal event of $H \to WW^* \to jj\ell\nu$, 
a Higgs particle near its mass shell (in the 125 GeV region) will
decay into two $W$ bosons. One of these will be essentially on-shell 
while the other will be highly virtual, with an invariant mass roughly 
equal to $40$ GeV. Either the hadronically or leptonically decaying 
$W$ may be the on-shell particle and the events are approximately evenly 
distributed between these two cases. 

The dominant physics background for our signal is $Wjj$ where the jets 
are produced by QCD processes. Thus, before any selection cuts, the
background is typically an on-shell leptonically decaying $W$ and a
pair of jets which can fake a second $W$ whether real or virtual. 
The dijet invariant mass distribution is, to first order, a smoothly 
falling function for the QCD background. Hence, to minimize background 
we select events with a higher dijet invariant mass and a
leptonic invariant mass which is far from the on-shell $W$ mass. 
Therefore we will concentrate on the half of the signal with an
on-shell $W$ decaying into two jets and a virtual leptonically decaying $W$.

The neutrino momentum is not directly measured so we must make some
assumptions to reconstruct the leptonic $W$ or the Higgs invariant
mass. Previous analysis have often used the assumption that 
the neutrino comes from an on-shell $W$, which is not suitable 
for our case. We will assume that the transverse neutrino momentum 
can be approximated by the missing transverse energy ($\notE_T$), 
computed from the sum of all detected particles. 
The Higgs invariant mass can be approximately located by using 
the cluster transverse mass~\cite{Han:1998ma}, defined as 
\begin{eqnarray}
M_C \equiv \sqrt{M_{jj\ell}^2 +\notE_T^2} +\notE_T
\end{eqnarray}
where $M_{jj\ell}$ is the invariant mass of the 2-jet plus charged
lepton system.

This quantity can be understood as the invariant mass constructed 
from the known momenta (assuming $\notE_T$ for the transverse neutrino 
momentum) with the longitudinal neutrino momentum chosen so as to minimize it. 
Equivalently, it corresponds to the invariant mass at an endpoint in the 
physically allowed parameter space with real momenta. 
We will use this principle to reconstruct the neutrino's longitudinal 
momentum as detailed in Section III. 

The cluster transverse mass is particularly useful in this scenario 
because of the low Higgs mass in comparison to the real $WW$ diboson mass. 
For the signal, the actual invariant mass of the $jj\ell\nu$ system 
is typically near the minimum value allowed by the visible particles 
plus $\notE_T$. As we raise cuts on the energy of the jets or 
the leptons, so long as we do not move beyond the range of energies 
the signal can produce, the $jj\ell\nu$ invariant mass will still be 
at the relatively low mass, typical of the Higgs resonance, and $M_C$ 
will be a good approximation to the actual mass. 
For the background, higher cuts on the produced particles will favor 
a higher $M_C$ since there is no resonance which keeps a low minimum 
invariant mass as a physical solution.  

The signal also has a characteristic spin structure which we consider
as a potential discriminant against background~\cite{Dobrescu:2009zf}. 
In our Higgs signal, the Higgs boson is a scalar decaying into 
vector bosons with opposite spins and those $W$ bosons couple 
only to left-handed particles in their decays. 
As a result, the up-type quark coming from the decay of one $W$ will 
tend to be aligned with the charged lepton coming from the other, 
while the down-type quark will tend to be aligned with the neutrino. 
In general we do not know which jet originates from the up-type quark, 
but we can still try to select for events where one jet is aligned 
and the other anti-aligned with respect to the  charged lepton. 

This phenomenon can be characterized by the angles $\phi$, $\theta_j$ 
and $\theta_l$. $\phi$ is defined as the angle between the $\ell\nu$ 
and the $jj$ decay planes in the rest frame of the Higgs. 
$\theta_j$ is the angle in the rest frame of the hadronically 
decaying $W$ between the leading jet (in energy) and the direction of 
boost from the Higgs rest frame.  
$\theta_\ell$ is similarly defined, with the charged lepton in place
of the leading jet. The signal is maximized for $\phi \simeq 0, \pi$ and 
for $\theta_j, \theta_\ell \simeq \frac{\pi}{2}$. 

\section{Signal Reconstruction}

As discussed above, we can use the cluster transverse mass $M_C(H)$ 
to approximate the resonance peak of the Higgs. Assuming the neutrino 
transverse momentum $k_T$ can be identified with the missing
transverse energy for the event, this is equivalent to choosing 
the longitudinal momentum of the neutrino as 
\begin{eqnarray}
  k_z = \frac{p^{vis}_z k_T}{\sqrt{(E^{vis})^2 - (p^{vis}_z)^2}}
\end{eqnarray}
where $E^{vis}$ and $p^{vis}$ are the energy and 3-momentum of the sum 
over the three visible particles $j$, $j'$ and $\ell$.

The same concept can be applied to the transverse mass 
$M_T(W) = M_T(\ell,\notE_T)$ often used with leptonically decaying $W$'s: 
\begin{eqnarray}
M_T(W)^2 & \equiv &
 (E^\ell_T + E^\nu_T)^2 - (\vec{p}^\ell_T +\vec{k}_T)^2 \, ,\\
k_z & = & \frac{p^\ell_z k_T}{E^\ell_T} \, ,  
\end{eqnarray}
where
\begin{eqnarray}
\vec{k}_T = \vec{\notE}_T \quad {\rm and} \quad E^\nu_T = \notE_T \, .
\end{eqnarray}
This method of assigning neutrino momentum is essentially the same as 
in the modified MAOS (MT2 Assisted On-Shell) method detailed in 
Ref.~\cite{Choi:2009hn} for use with two invisible particles.

Since we expect a low invariant mass ($M_{\ell\nu}$) for the virtual $W$ 
from Higgs decay, $M_T(W)$ and its associated $k_z$ value can also
work as signal discriminant. 
Let us call the longitudinal momentum of the neutrino in the first
scheme above $k_z(H)$ and in the second $k_z(W)$. 
In general $k_z(H)$ will perform slightly better for reconstructing 
the Higgs mass near its true peak and $k_z(W)$ is slightly better for $W$ 
reconstruction, particularly at higher values of $M_{jj}$.

We have also considered an intermediate, weighted case using the prescription 
\begin{align}
K_z =
 \frac{ [p^{vis}_z M_T(W)^2 +p^\ell_z M_C(H)^2]k_T }
{ \sqrt{ [E^{vis} M_T(W)^2 +E^\ell M_C(H)^2]^2
 -[p^{viz}_z M_T(W)^2 +p^\ell_z M_C(H)^2]^2}}
\end{align}
which approximately minimizes the product $M_H \times M_W$ when used in 
reconstructions. In practice, after cuts to select the mass peaks, 
there are only small differences in the distributions resulting from 
using $k_z(H), k_z(W)$ or $K_z$. 
In our analysis, we assign the longitudinal momentum of the neutrino 
according to $k_z(H)$, which appears to give us slightly better 
signal discrimination than the other options. 
$k_z(H)$ gives the sharpest edge to the Higgs mass peak and also 
performs well for the $W$ reconstruction in the far below shell 
region we are selecting.

\section{Event Simulations}

We perform Monte Carlo simulations for the signal and the background 
events using the MadGraph5 package~\cite{Alwall:2011uj}. 
Our typical signal jets do not have particularly high momentum so 
we are sensitive to contamination from initial-state radiation. 
To control this, and to have a better estimate of signal and
background shapes, we use the built-in MLM-style matching scheme. 
This option combines matrix element and showering routines 
in a consistent way to avoid over-counting. 
We include up to one additional jet at the matrix element level 
in both signal and background. Showering and hadronization is
performed by the event generator PYTHIA~\cite{Sjostrand:2006za}, 
after which our events are passed to the Delphes fast detector
simulation for reconstruction~\cite{Ovyn:2009tx}. 
At the Delphes level, we define our jets according to the 
Cambridge-Aachen (C-A) algorithm with a size parameter of 
$\Delta R  \equiv \sqrt{\Delta \phi^2 + \Delta \eta^2 }= 0.5$. 

We require at least one isolated lepton ($\ell = e$ or $\mu$) 
in each event and take the leading lepton in transverse momentum 
($p_T$) as our candidate from the leptonic $W$ decay. 
Since Delphes includes electrons in its listing of jets, 
we subtract the lepton momentum from any jet within 
a $0.5$ cone in $\Delta R$ and recombine any remaining momentum 
according to the C-A prescription. 
The transverse momenta of our jets are typically $\sim 40$ GeV 
or less. Energy loss from hadronization, reconstruction and detector
effects can be significant for jets in this momentum range.
To ameliorate this we apply a jet-energy 
correction factor according to the pseudo-rapidity and magnitude of 
the momentum of each candidate jet. This correction factor is based on 
comparison between jets at reconstruction level and quarks/gluons 
at parton level  when they can be well matched, averaged over 
a large number of background and signal simulated events. 
For jets with momentum $|\vec{p}| \lesssim 20$ GeV this can be 
an order one correction. We apply a similar correction procedure 
to the charged lepton, although that is only a small adjustment.

As noted above, the background is dominated by $Wjj$ production. 
We separate this into two pieces, a leading QCD piece with only 
two electroweak vertices, and a sub-leading piece with 
four electroweak vertices which includes non-Higgs-generated $W^+W^-$ 
events. We also consider $t\bar{t}$ events. 

The Higgs signal is produced primarily through gluon fusion, which is 
implemented in MadGraph via an effective theory derived from one loop 
calculations with the top quark. 
However, the total production is significantly enhanced 
at higher order, suggesting a K-factor of $\sim 2$ compared to our
leader order (LO) simulations. 
To take this into account we scale our signal results for 
$pp \to H +X$ up to match the higher order (NNLO) results, which find 
a production cross-section of $19.5$ pb at 8 TeV and $49.8$ pb 
at 14 TeV~\cite{Dittmaier:2011ti}. 
We provide results for a 123, 125 and 127 GeV Higgs in Table I. 
For the 123 GeV and 127 GeV cases we assume the same scaling as 
for $M_H = 125$ GeV. 

For the backgrounds, we have made use of the MCFM program suite for 
computing $Wjj$ at the next-to-leading-order (NLO)~\cite{Campbell:2010ff}. 
We impose a $p_T$ cut of 5 or 10 GeV and require an invariant mass cut 
55 GeV $< M_{jj} <$ 105 GeV for the NLO results. 
We impose the same mass cut for our MadGraph LO plus matching
simulation. The matching algorithm has an implicit cutoff $p_T \sim 10$ GeV 
which defines the boundary between matrix element and showering effects. 
At 7 TeV the NLO and LO+matching cross-sections agree quite well and 
are stable when varying the $p_T$ cut between 5 and 10 GeV.  
At 14 TeV the NLO estimates are approximately $15 \%$ higher than 
LO + matching, although with estimated errors of the same order. 
For the results presented below we do not apply a K-factor beyond our 
LO + matching calculations for our $Wjj$ backgrounds. For the 
$t\bar{t}$ background we include a K-factor of 2. 

Figure~1 shows invariant mass distributions ($d\sigma/dM_{jj\ell\nu}$) 
with basic cuts: $p_T(j) \ge 5$ GeV, $p_T(\ell) \ge 20$ GeV, 
$|\eta(j)| \le 5$, and 55 GeV $< M_{jj} <$ 105 GeV.
In each event, we assume that there are two jets and one isolated leptons 
as well as missing transverse energy from a neutrino.
In this figure, we present the reconstructed masses for the signal 
with $M_H = 125$ GeV and for the background from $Wjj$.


\begin{figure}[htb]
\centering\leavevmode
\includegraphics[angle=270,scale=0.5]{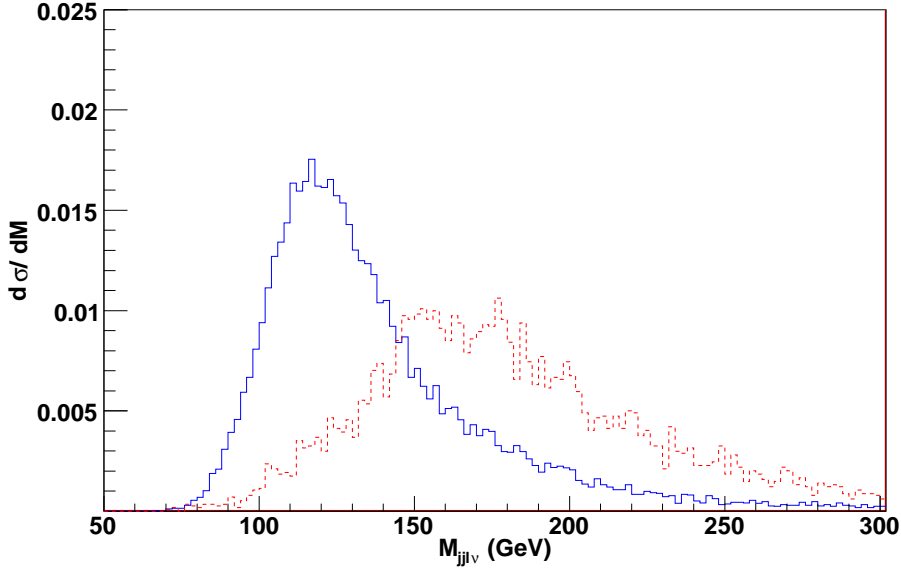}
\caption[]{
Invariant mass distribution of $jj\ell\nu$ with basic cuts on
$p_T(j)$, $\eta(j)$, and $M_{jj}$ for
(a) the Higgs signal from $pp \to H \to WW^* \to jj\ell\nu +X$ 
with $M_H =$ 125 GeV (blue solid), and 
(b) the dominant physics background (red dash) from 
$pp \to Wjj \to \ell\nu jj +X$. 
We have normalized the cross section for both the signal and the background.
\label{fig:hww_mass} }
\end{figure}

\section{Acceptance Cuts}

We apply a series of acceptance cuts to improve the statistical 
significance. We first require that all events have at least two jets
and one isolated charged lepton.  After jet-energy corrections, 
the first and second leading jets by transverse momentum are required 
to have $p_T(j_1) > 30$ GeV and $p_T(j_2) > 20$ GeV. 
The invariant mass of this jet pair ($M_{jj}$) must be between $65$ 
and $95$ GeV. Conversely, the charged lepton must have a transverse momentum 
$p_T(\ell) < 30$ GeV, and the missing transverse energy $\notE_T$ can
be capped at $40$ GeV. We consider jets with a pseudo-rapidity 
$|\eta_j| < 5$ and require the charged lepton to have $|\eta_\ell| < 2.5$.

With these inputs we reconstruct the longitudinal neutrino momentum 
as described above and equate the neutrino transverse momentum to
$\notE_T$. Using this assumption we can calculate the momentum of 
$W_{l\nu}$, the leptonically decaying weak boson, and $H$, 
the candidate Higgs boson. 

In addition, we impose the following cuts:
\begin{itemize}
\item $M_{l\nu} < 45$ GeV,
\item $M_H \simeq M_{jj\ell\nu} < 130$ GeV,
\item $\Delta R_{j\ell} > 0.2$, and 
\item $E^0_{\ell\nu} < 45$ GeV, 
\end{itemize}
where $E^0_{l\nu}$ is the energy of the leptonically decaying $W$ 
in its rest frame. With these cuts applied, the remaining background 
is kinematically similar to the signal, although the signal's 
characteristic peaks are somewhat sharper. Further tightening the cuts 
can reduce the ratio of signal to background but generally reduces 
the statistical significance due to loss of signal. 
We do not find that angular correlations in the variables 
$\phi$, $\theta_j$ and $\theta_l$ are sufficiently distinct 
from the background to improve our results.

\section{Discovery Potential at the LHC}

With the procedures and cuts discussed above, we present our estimates
of signal and background rates in Table I. We consider two cases: 
the LHC 2012 running at $8$ TeV CM energy, and the planned LHC 
running at a target CM energy of 14 TeV. For 14 TeV we raise the 
$p_T$ cut on the second jet to $25$ GeV. The results include a 
signal calculated for input Higgs masses of 123, 125, and 127 GeV. 
We use the same cuts and one can see in Table I that the difference 
in expected signal events is small, although slightly increasing for
higher masses. This is owing to the increasing $WW$ branching fraction 
as the Higgs mass increases, an effect which it mitigated by the decreasing 
efficiency of the $M_{jj\ell\nu} < 130$ GeV cut as the signal peak moves
up in mass. Obviously, this cut  would drastically reduce our signal 
for masses much larger than those considered.



\begin{table}[h]
\label{sigma1}
\caption[]{Cross section of $pp \to H \to WW^* \to jj\ell\nu +X$ in fb 
at the LHC with all acceptance cuts for three values of 
$M_H = 123, 125,$ and 127 GeV and two values of CM energy (a)
$\sqrt{s} =$ 8 TeV and (b) $\sqrt{s} =$ 14 TeV.  
Also shown are the contributions from dominant physics backgrounds and 
the statistical significance for the Higgs signal with an integrated luminosity of 25 fb$^{-1}$.}
\begin{tabular}{|c|c|c|c|c|c|}
\hline
$\sqrt{s}$ & Signal(123/125/127) & $Wjj$(QCD) & $Wjj$(EW)
 & $t\bar{t}$ & $N_{SS} \equiv S/\sqrt{B}$ \\
\hline
(a) 8 TeV & 102/105/106 & 6170 & 93.5 & 19.0 & 6.4/6.6/6.6 \\
\hline
(b) 14 TeV & 175/188/201 & 8170 & 134 & 46.7 & 9.5/10./11. \\
\hline
\end{tabular}
\end{table}

We assume that the $8$ TeV running will accumulate an integrated 
luminosity ($L$) of $25 \, \text {fb}^{-1}$ for each detector 
of ATLAS or CMS. The statistical significance is defined as 
$N_{SS} \equiv S /\sqrt{B}$, where $S = L\times \sigma_S$ is the
number of signal events, $B = L\times \sigma_B$ is the number of 
background events, and $\sigma_{S,B}$ is the cross section of 
the signal or the background. 
Based on our numbers above this would give a statistical 
significance of $6.6 \sigma$ for the 2012 run in this channel. 
A combined analysis of the data from both CMS and ATLAS could
therefore potentially approach $9 \sigma$. 
At $14$ TeV a $5 \sigma$ discovery could be made with 
$L = 6 \, {\text fb}^{-1}$ for a single detector, not including any
data from 2012 running. 
We should stress at this point that our signal to background ratio 
is small, on the order of $1-2 \%$. Thus systematic uncertainties 
on the expected size of the background become very important and 
may wash out the purely statistical significance quoted above. 
Nonetheless, the signal features a distinct kinematic feature 
in the reconstructed Higgs peak near $125$ GeV, so that measurements 
of the background outside the peaked area can help constrain 
the true background.

\section{Prospects with c-tagging}

In this section we consider a proposal advanced by Sullivan and Menon
to study this channel with dedicated c-tagging algorithms. 
Many searches make use of b-tagging algorithms to better discriminate 
signal from background, and top-tagging programs have also 
been proposed~\cite{btagging,toptagging}. 
At present, there are no procedures specifically designed to 
distinguish c-quark jets from light quarks and gluons. 
In practice, b-tagging, sometimes referred to as heavy-flavor tagging, 
already has some utility for this purpose. Jets arising from c-quarks 
are mis-tagged as b-quark jets at a higher rate than those arising
from lighter quarks and gluons. 
Let us consider $\eps_b$ as the b-tagging efficiency, $\eps_c$ being
the effective rate of a $c$-jet mis-tagged as a b-jet, and 
$\eps_j$ is the mis-tagging rate for $u,d,s,g$-jets. 
The ratio of $\eps_b /\eps_j$, is an acceptance parameter that
characterizes the `tightness' of the b-tagging algorithm. 
At the ATLAS or the CMS~\cite{ATLAS,CMS}, 
for a b-tagging efficiency of approximately $\eps_b \sim 50-60\%$, 
the c-mistag rate is $\eps_c \sim 10-15 \%$, 
while the light-jet mistag rate is $\eps_j \lesssim 1 \% $. 
Thus, the principle of a dedicated c-tagger is plausible although 
it remains to be developed.

For the discovery channel explored in this letter, c-tagging provides
two advantages. The first is that half the events in our signal should 
involve a $W$ decaying to a charm quark ($c$) and a strange quark ($s$), 
and are thus amenable to c-tagging. In contrast, our backgrounds are 
dominated by light jets with only $\sim 1/6$ of the events involving 
a final state $c$-jet. The second is that tagging the $c$-jet in 
our signal allows us to better use the angular correlations discussed
above. Without tagging we do not know which jet arises from the
$u$-quark or the $c$-quark, and can only say that one jet or the other 
should be correlated/anticorrelated with the charged lepton
direction. C-tagging would resolve this ambiguity and increase 
the usefulness of angular correlations as an experimental discriminant. 

The requirement of c-tagging necessarily suppresses our overall signal 
rate, and this reduction in statistics might hurt our significance. 
Therefore any c-tagging scheme would need to be highly efficient 
to preserve our signal acceptance, while still rejecting most light
jets. Using current b-tagging algorithms as a model, a high c-tagging 
acceptance can be achieved simply by raising the b-tagging
acceptance. This is not a problem for our signal since, even with 
$100 \%$ acceptance of b-jets, they would constitute only a small 
fraction of our backgrounds. However, for current b-tagging
algorithms, a high acceptance reduces the ratio of $c$-(mis)tag 
to $udsg$-mistag rates. As will be seen below, a successful
application of c-tagging to this signal would require high $c$-jet 
acceptance with better light jet rejection than appears possible 
with the existing algorithms.

For the analysis with possible $c$-tagging, we include the same
backgrounds as before. In addition, we divide the total $Wjj$
backgrounds into those including at least one $c$ or $\bar{c}$ 
at the parton level ($Wcj$) and those including only light partons 
(labeled $Wjj$). We perform the same reconstruction and cuts as 
described above, with the following modifications: 
We require at least one c-tagged jet and we consider the leading 
c-tagged jet in $p_T$ as our candidate $c$ quark from $W$ decay. 
For the second jet we use the leading non-tagged jet or the
second-leading c-tagged jet if one is present, whichever is higher 
in $p_T$. Reconstruction of $W$s and the Higgs is performed as above. 
We apply $p_T$ cuts on the two chosen jets of $p_T > 30,25$ GeV 
on the first and second jet ordered by $p_T$. 
All other cuts from the untagged analysis are the same. 
Additionally, we apply the following cuts on the angular variables:
\begin{itemize}
\item $\phi > 1.2$ radians,
\item $ (0.9 \cos \theta_l -1.2) < \cos \theta_c < (1.1 \cos \theta_l +1)$.
\end{itemize}
Here $\theta_c$ is the angle between the $c$-jet and the boost
direction of the hadronic decaying $W$ in the rest frame of the $W$, 
rather than the angle for the leading $p_T$ jet as used in the untagged case.

The cross sections of the signal and the background with $c$-tagging
are given in Table II. Each sub-channel must be multiplied by an
effective tagging efficiency for a hypothetical or existing tagging
algorithm as indicated. Note that while $\eps_c$ is essentially the
single-jet c-tagging efficiency, with a small enhancement coming from 
mis-tagged light jets in $Wcj$ channels, $\eps_j$ should include the 
probability of mis-tagging any light jet in a $Wjj$ channel. 
For the backgrounds, which will sometimes include additional jets
after showering and reconstruction, 
we will use $\eps_j^{eff} = 2.5 \epsilon_j^0$ where $\eps_j^0$ is 
the single light jet mistag rate. 
For the $t\bar{t}$ background, $b$ quarks from top decay are likely to
be tagged as $c$-jets. In our estimates we will assume that for a 
high acceptance c-tagger every $t\bar{t}$ event will have at least 
one tagged $c$-jet.


\begin{table}[h]
\label{sigma2}
\caption[]{Cross section of $pp \to H \to WW^* \to jj\ell\nu +X$ in fb 
at the LHC with all acceptance cuts and c-tagging for 
$M_H =$ 125 GeV and two values of CM energy: (a) 8 TeV and (b) 14 TeV.  
Also shown are the contributions from dominant physics backgrounds.}
\begin{tabular}{|c|c|c|c|c|c|c|}
\hline
$\sqrt{s}$ 
  & Signal & $Wcj$(QCD) & $Wjj$(QCD) & $Wcj$(EW) & $Wjj$(EW) & $t\bar{t}$ \\
\hline
(a) 8 TeV & 28.3$\epsilon_c$ & 162.2$\epsilon_c$ & 904.6$\epsilon_j$ & 13.6$\epsilon_c$ &12.6$\epsilon_j$ & 9.16$\epsilon_b$ \\
\hline
(b) 14 TeV & 56.6$\epsilon_c$  &472.3$\epsilon_c$  & 1734.$\epsilon_j$ &  21.3$\epsilon_c$ & 22.1$\epsilon_j$ & 22.7$\epsilon_b$ \\
\hline
\end{tabular}
\end{table}

In the table above one can see that c-tagging does potentially improve
our statistical significance, as well as improving the ratio of signal 
to background. However, realizing this potential would require
excellent c-tagging acceptance while keeping the ratio $\eps_j/\eps_c$
low. 
In the ideal case, where $\eps_c \simeq 1$ and $\eps_j \simeq 0.01$, 
we would have $9.7 \sigma$ at $\sqrt{s} = 8$ TeV based on 
statistical uncertainty. At $\sqrt{s} = 14$ TeV a $5 \sigma$ detection 
could be made with $L = 4.5 \; \text{fb}^{-1} $. On the other hand, 
let us consider the more modest but still optimistic case where
c-tagging has a similar performance to current b-tagging. 
If $\epsilon_c = 0.5$ with $\epsilon_j^0 = 0.01$, our nominal
significance with 2012 data would be $6.5 \sigma$, virtually the same
as the untagged case. However, the signal to background ratio would be 
improved to $\sim 10 \%$. Thus an efficient c-tagging can reduce our 
sensitivity to background systematics. At $\sqrt{s} = 14$ TeV the
statistical significance would be somewhat worse than the untagged
case, requiring $L = 9.8 \, \text{fb}^{-1} $ for a $5 \sigma$ result. 
This is because the background after c-tagged cuts, especially the
$Wcj$ component, grows more quickly with increasing beam energy than 
the overall background with untagged cuts.

\section{Conclusions} 

In this letter we have investigated the discovery channel 
$H \to WW^* \to jj\ell\nu$ for a Standard Model Higgs with a mass of
125~GeV, consistent with recent LHC results. We have demonstrated that 
by selecting for an on-shell hadronically decaying $W$ paired with 
a far-off-shell leptonic decaying $W$, combined with transverse-mass 
based reconstruction techniques, one can reduce the large $Wjj$ 
backgrounds to a workable level. Based on Monte Carlo simulations 
we estimate that the 2012 run of the LHC could provide evidence for 
this channel at the $6\sigma$ level with an integrated luminosity of 
25~fb$^{-1}$ based on statistical uncertainty. 
At the design energy of 14 TeV,  $5\sigma$ significance could be
achieved with $L = 6 \, \text{fb}^{-1}$ of data. However, this
analysis does not include a full estimation of systematic
uncertainties which will play an important roll given the small ratio 
of signal to background. Careful study of the $Wjj$ background will be 
required to make this channel feasible. 
Nonetheless, our results are promising. 

We also considered the prospects for c-tagging to improve our
results. We find that exceptional c-tagging capabilities, with 
high acceptance and good rejection of light jets, could yield somewhat 
improved statistical significance. However, with more realistic
assumptions for c-tagging efficiencies, we would have at best marginal 
improvement in terms of significance. On the other hand, the increased 
signal to background ratio is a distinct advantage of this scenario.

We note that our study is based on a simulation of traditional
calorimeter-based jets. Due to the relatively low energy of our
typical jets, we are quite sensitive to loss of resolution from 
energy loss and uncertainty in jet-energy corrections. 
This limits our ability to pick out the pronounced hadronic decaying 
$W$ mass peak and the Higgs transverse mass peak. A study with
particle-flow based jet reconstruction may well be able to improve 
on our findings.

\section*{Acknowledgments}

We are grateful to Bill Kilgore for beneficial discussions.
This research was supported in part by the U.S. Department of Energy
under Grant No.~DE-FG02-04ER41305.

\newpage



\begin{thebibliography}{20}

\bibitem{atlashiggs} 
  G.~Aad {\it et al.}  [ATLAS Collaboration],
  Phys.\ Lett.\ B {\bf 716}, 1 (2012)
  [arXiv:1207.7214 [hep-ex]].

\bibitem{cmshiggs} 
  S.~Chatrchyan {\it et al.}  [CMS Collaboration],
  Phys.\ Lett.\ B {\bf 716}, 30 (2012)
  [arXiv:1207.7235 [hep-ex]].

\bibitem{atlasllvv} 
  G.~Aad {\it et al.}  [ATLAS Collaboration],
  Phys.\ Lett.\ B {\bf 716}, 62 (2012)
  [arXiv:1206.0756 [hep-ex]].

\bibitem{cmsllvv} 
  [CMS Collaboration],
  CMS-PAS-HIG-12-038;
  S.~Chatrchyan {\it et al.}  [CMS Collaboration],
  Phys.\ Lett.\ B {\bf 710}, 91 (2012)
  [arXiv:1202.1489 [hep-ex]].

\bibitem{Barate:2003sz} 
  R.~Barate {\it et al.}  [LEP Working Group for Higgs boson searches
    and ALEPH and DELPHI and L3 and OPAL Collaborations],
  Phys.\ Lett.\ B {\bf 565}, 61 (2003)
  [hep-ex/0306033].

\bibitem{Han:1998ma} 
  T.~Han and R.~-J.~Zhang,
  Phys.\ Rev.\ Lett.\  {\bf 82}, 25 (1999)
  [hep-ph/9807424].

\bibitem{Lykken:2011uv} 
  J.~D.~Lykken, A.~O.~Martin and J.~-C.~Winter,
  arXiv:1111.2881 [hep-ph].

\bibitem{atlasjjlv} 
  G.~Aad {\it et al.}  [ATLAS Collaboration], 2012 
  arXiv:1206.6074 [hep-ex].

\bibitem{cmsjjlv} 
  CMS Collaboration, 2012 
  CMS-PAS-HIG-12-021.

\bibitem{Menon:2010vm} 
  A.~Menon and Z.~Sullivan,
  arXiv:1006.1078 [hep-ph].

\bibitem{Dobrescu:2009zf} 
  B.~A.~Dobrescu and J.~D.~Lykken,
  JHEP {\bf 1004}, 083 (2010)
  [arXiv:0912.3543 [hep-ph]].

\bibitem{Choi:2009hn} 
  K.~Choi, S.~Choi, J.~S.~Lee and C.~B.~Park,
  Phys.\ Rev.\ D {\bf 80}, 073010 (2009)
  [arXiv:0908.0079 [hep-ph]].

\bibitem{Alwall:2011uj} 
  J.~Alwall, M.~Herquet, F.~Maltoni, O.~Mattelaer and T.~Stelzer,
  JHEP {\bf 1106}, 128 (2011)
  [arXiv:1106.0522 [hep-ph]].

\bibitem{Sjostrand:2006za} 
  T.~Sjostrand, S.~Mrenna and P.~Z.~Skands,
  JHEP {\bf 0605}, 026 (2006)
  [hep-ph/0603175].

\bibitem{Ovyn:2009tx} 
  S.~Ovyn, X.~Rouby and V.~Lemaitre,
  arXiv:0903.2225 [hep-ph].

\bibitem{Dittmaier:2011ti} 
  S.~Dittmaier {\it et al.}  [LHC Higgs Cross Section Working Group Collaboration],
  arXiv:1101.0593 [hep-ph].

\bibitem{Campbell:2010ff} 
  J.~M.~Campbell and R.~K.~Ellis,
  Nucl.\ Phys.\ Proc.\ Suppl.\  {\bf 205-206}, 10 (2010)
  [arXiv:1007.3492 [hep-ph]].

\bibitem{btagging} 
  [CMS Collaboration],
  CMS-PAS-BTV-11-004.

\bibitem{toptagging} 
  T.~Plehn and M.~Spannowsky,
  arXiv:1112.4441 [hep-ph].

\bibitem{ATLAS}
ATLAS Collaboration, ``ATLAS Detector and Physics Performance Technical
Design Report, CERN/LHCC 99-14/15 (1999)'';
  G.~Aad {\it et al.}  [The ATLAS Collaboration],
  ``Expected Performance of the ATLAS Experiment - Detector, Trigger and
  Physics,'' (2009).

\bibitem{CMS}
  G.~L.~Bayatian {\it et al.}  [CMS Collaboration],
  ``CMS technical design report, volume II: Physics performance,''
  J.\ Phys.\ G {\bf 34}, 995 (2007).

\end{thebibliography}
\end{document}